\documentclass[superscriptaddress,showpacs,twocolumn,prb]{revtex4}
\usepackage{epsfig,amssymb,amsmath,graphicx}
\begin{document}
\title{Manifestation of Resonance-Related Chaos in Coupled Josephson Junctions}
\author{Yu. M. Shukrinov}
\affiliation{BLTP, JINR, Dubna, Moscow Region, 141980, Russia}
\author{M. Hamdipour}
\affiliation{BLTP, JINR, Dubna, Moscow Region, 141980, Russia}
\affiliation{Institute for Advanced Studies in Basic Sciences, P.O. Box 45195-1159, Zanjan, Iran}
\author{M. R. Kolahchi}
\affiliation{Institute for Advanced Studies in Basic Sciences, P.O. Box 45195-1159, Zanjan, Iran}
\author{A. E. Botha}
\affiliation{Department of Physics, University of South Africa, P.O. Box 392, Pretoria 0003, South Africa}
\author{M. Suzuki}
\affiliation{Photonics and Electronics Science and Engineering Center and Department of Electronic Science and Engineering,  Kyoto University, Kyoto 615-8510, Japan}
\date{\today}
\begin{abstract}
Chaotic features of systems of coupled Josephson junctions are
studied. Manifestation of chaos in the temporal dependence of the
electric charge, related to a parametric resonance, is demonstrated
through the calculation of the maximal Lyapunov exponent, phase-charge
and charge-charge Lissajous diagrams and correlation functions.
The number of junctions in the stack strongly influences the fine structure
in the current voltage characteristics and a strong proximity effect
results from the nonperiodic boundary conditions. The observed
resonance-related chaos exhibits intermittency over a range of
conditions and parameters. General features of the system are analyzed
by means of a linearized equation and the criteria for a breakpoint
region with no chaos are obtained. Such criteria could clarify recent
experimental observations of variations in the power output from
intrinsic Josephson junctions in high temperature superconductors.
\end{abstract}
\maketitle
\date{}
\section{Introduction}
Systems of coupled Josephson junctions (JJs) are prospective objects
for superconducting electronics and therefore continue to be the
subject of intense
investigations.\cite{kleiner92,kadowaki,krasnov11,savelev10,ozyuzer,tachiki09,wang10,lin11,koshelev10} Intrinsic
Josephson effects -- tunneling of Cooper pairs between superconducting
layers inside of strongly anisotropic layered high temperature
superconductors (HTSCs) -- provide an opportunity to model HTSCs as
systems of coupled intrinsic Josephson junctions (IJJs) and an
effective way to investigate nonlinear effects and non-equilibrium phenomena
in HTSCs.  Intrinsic
tunneling is a powerful method for the study of the nature of high
temperature superconductivity, transport along a stack of
superconducting layers, and the physics of vortices. It also plays
an important role in determining the
current-voltage characteristics (CVCs) of tunneling structures based
on HTSCs and the properties of vortex structures in these
materials.

Parametric resonance in coupled JJs demonstrate a series of novel
effects predicted by numerical simulations.\cite{sg-prb11} In
comparison with the single junction, the system of coupled JJs has a
multiple branch structure in its CVCs.  The outermost branch of the
CVCs has a breakpoint (BP) and a breakpoint region (BPR) before a
transition to another branch, caused by parametric
resonance.\cite{sm-prl07} The BP current characterizes the resonance
point at which a longitudinal plasma wave (LPW) is created. It was
demonstrated that the CVC of the stack exhibits a fine structure in
the BPR.\cite{sms-prb08} The breakpoint features were predicted
theoretically\cite{sm-sust07}  and recently observed
experimentally.\cite{iso-apl08} A breakpoint region also appears in
numerical simulations performed by other authors.\cite{machida99}

The physics of IJJs, which are nonlinear systems, cannot be
completely understood without the investigation of their chaotic features.
Parametric instabilities are common phenomena in one-dimensional arrays of
Josephson junctions. For example, a one-dimensional parallel array of $N$
identical Josephson junctions was studied by discrete sine-Gordon equation
(also known as Frenkel-Kontorova model) in Ref.\onlinecite{watanabe95}. The parametric
instabilities were predicted by theoretical analysis and observed
experimentally. In particular, the novel resonant steps related to the
parametric instability were found in the current voltage characteristics of
discrete Josephson ring. It was verified experimentally that such steps occur
even if there were no vortices in the ring. A long Josephson junction is also
often considered as a one-dimensional parallel array of $N$ identical
Josephson junctions. An experimental testing of the parametric instability in
IJJ would clarify the generic feature of one-dimensional parallel and series
array models and the role played by resonance related chaos in these
systems.

The chaotic features of IJJs in HTSCs are also of great
interest due to the observed powerful coherent THz radiation from such
systems.\cite{ozyuzer}
Broadly tunable sub-terahertz emission
was found in the CVC near the switching point from the outermost branch to inner
branches, and also from inner branches.\cite{kadowaki}
We stress that this is the same region
in the CVCs where the BP and BPR were observed. Recently, for example,
coherent THz radiation has been measured experimentally and was
associated with a ``bump'' structure corresponding to the same part of
the CVCs.\cite{ozyuzer12,benseman} Obviously, the chaos in IJJs could
strongly effect their radiation properties, because the transition to
the chaotic state may inhibit the coherent radiation from
IJJs. Therefore, detailed investigations of such transitions in real
systems are very important. This situation makes the phase dynamics
and the investigation of chaos in intrinsic JJs, near the BPR
of the CVCs, an urgent problem.

In an early study of the rf current driven Josephson junction it was
found that, depending on the amplitude of the rf current, the system
would develop a chaotic
character.\cite{Huberman80,Kautz81,Kautz85a,Kautz85b} It was also
shown that the onset of chaos as a function of frequency correlated
with the function that indicated infinite gain, also as a function of
frequency, for the unbiased parametric amplifier.\cite{pedersen} More
recently the importance of chaos in IJJs, and its effects on the CVCs
and the Shapiro steps in these systems, was stressed in
Refs. \onlinecite{irie03,scherbel04}.

In a further comprehensive study, the critical behavior of the
dynamical equation, in the sense of how it goes from regular to
chaotic, was investigated.\cite{tomas} It was discovered that the
critical behavior is that of the circle map. Here, in the subcritical
state, the resonances are separated by quasi-periodic orbits
(i.e. having irrational winding number), whereas in the supercritical
state the dynamics is composed of chaotic jumps between
resonances. The chaos appears to be the result of the overlap of
resonances.\cite{mogens}

In a similar study, the interaction of the Josephson junction with its
surroundings was considered by coupling it in parallel with an RLC
circuit, modeling a resonant cavity.\cite{borcherds} The chaotic
nature develops through the familiar period doubling sequence of
bifurcations. Other studies exist that try to understand the irregular
response of dissipative systems driven by external sources when such
behavior is interleaved by the synchronized motion (Shapiro
steps). This so called intermittent chaos is modeled by a random walk
between the two neighboring steps that have become unstable. It is
shown that the power spectrum of the voltage correlations has a
broadband background characterizing the chaotic
solution.\cite{benjacob} It may be noted that a similar random walk
model has been developed to explain the phase synchronization of
chaotic rotators.\cite{osipov02}

Interesting features of chaos, chaos control and chaotic
synchronization in Josephson junction arrays (JJAs) and shunted JJ
models have been described in Refs.~\onlinecite{ri11,feng10,whan96,basler95,zhou09,hassel06,bhagavatula94,chernikov94}.
JJAs attract much attention  due to their perspective as high
precision voltage standards and, as mentioned earlier, high power
coherent THz sources. Output power of a single junction is extremely
low (typically 10 nW), but much higher output power can be obtained
by using JJAs. Chaos and nonlinearity in JJAs have been described in
Refs.~\onlinecite{basler95,zhou09}.  In particular, Basler {\em et
  al.}\cite{basler95} reported on the theory of phase locking and
self-synchronization in a small JJ cell by using the resistively
shunted junction (RSJ) model and Hassel {\em et al.}\cite{hassel06}
studied self-synchronization in distributed Josephson junction
arrays. Features of spatiotemporal chaos in JJAs composed of RSJs were
numerically investigated by Bhagavatula {\em et
  al.}\cite{bhagavatula94} Chernikov and Schmidt\cite{chernikov94}
demonstrated that when four or more JJs are globally coupled by a
common resistance, the system exhibits adiabatic chaos. Direct
calculation results \cite{zhou09} showed that phase locking and chaos
coexist when there are three junctions in the JJA. In
Ref.~\onlinecite{ri11} chaos, hyperchaos and controlling hyperchaos in
a JJA  were investigated. The numerical results show that chaos and
hyperchaos states can coexist in this array of two shunted JJs,
independently of whether or not the original states of the junctions
are chaotic.

Although there have been recent reports on the chaotic behavior
in JJs, such as Ref.~\onlinecite{ri11}, there are relatively few
detailed investigations of chaotic behavior in closely-related
phenomenological models such as, the capacitively-coupled model
(CCJJ),\cite{koyama96} the resistive-capacitive shunted model
(RCSJJ),\cite{buckel,whan96} or the capacitively coupled Josephson
junction model with diffusion current
(CCJJ+DC)\cite{sms-physC06,machida00} of the present work.
Specifically, the chaotic features at the parametric resonance
have not been investigated before.

In this paper, we study the phase dynamics in coupled Josephson
junctions. A resonance-type chaos related to the parametric resonance
in this system is demonstrated. The origin of chaos is related to the
coupling between junctions based on the fact that the superconducting
layers (S-layers) are in the nonstationary nonequilibrium state due to
the injection of quasiparticle and Cooper
pairs.\cite{koyama96,ryndyk-prl98} We present results on a chaotic
mode which  is due to the coupling between junctions and a parametric
resonance in coupled JJs. This mode cannot be observed in case of a
single JJ. Creation of a LPW and its interplay with a discrete lattice
of superconducting layers produces a rich dynamic behavior, including,
for example, intermittency. We use different methods to investigate
this dynamics, including correlation functions and Lyapunov exponents.
Results for stacks with different numbers of JJs, in which LPWs with
different wave numbers are exited, give interesting information on
certain features of the fine structure within the BPR of the CVCs.
The influence of the  boundary conditions on the chaotic part of the
fine structure leads to the manifestation of a strong proximity effect
in IJJs. Transitions between chaotic and regular behavior in BPR are
predicted. Transitions from the chaotic state to both a regular LPW
state and a traveling wave mode are demonstrated. Finally some general
features of the system are analyzed and the conditions and parameters
leading to a BPR with no chaos are found.

The paper is organized in the following way. In Section II we
introduce the coupled sine-Gordon equation, used for numerical
simulations, and discuss the boundary conditions and numerical
procedures. We briefly describe the method for calculating the
Lyapunov exponent. Section III is devoted to the study of the resonance-related
chaos produced in the temporal dependence of the
electric charge, and this is demonstrated
through the calculation of the maximal Lyapunov exponent, phase-charge
and charge-charge Lissajous diagrams and correlation functions.
We also provide examples of the manifestation of related
chaotic features in the correlation functions and polar diagrams. The effect of the number of
junctions in the stack is considered in Sec. IV. Nonperiodic boundary
conditions and the proximity effect in coupled JJs are discussed in
Sec. V. We show the occurrence of transitions from chaotic to regular behavior in Sec. VI. To clarify the chaotic features in coupled JJs, we analyze
the linearized equation for phase differences in Sec. VII, and discuss
the question concerning the region of instability at chaotic boundary. In Sec.  VIII we demonstrate the parametric resonance region
without chaotic behavior. Conclusions follow in Sec. IX.

\section{Model and method}
High-Tc superconductors have a layered crystal structure and strong
anisotropy. A model of multilayered Josephson junctions, which has an
alternating stack of superconducting and insulating layers, is sufficient to
describe the electronic properties of these systems\cite{matsumoto99}. In the present model we assume that the
physical quantities are spatially homogeneous on each layer. This assumption
is applicable in cases of no applied magnetic field and for a small sample
size in the $ab$-direction, i.e. for the intensively developing research field
known as the physics of Josephson nanojunctions, where the length of the
Josephson junctions is smaller than the Josephson penetration length.
One-dimensional models with coupling between junctions do capture the main
features of real IJJs, like hysteresis and branching of the current voltage
characteristics, and thus help to improve the understanding of their physics.
An interesting and very important fact is that the 1D models can also be used
to describe the properties of a parallel array of Josephson junctions, which
is often considered as a model for long Josephson junctions. The experiments
and theoretical studies on the propagation of Josephson fluxons and
electromagnetic waves in parallel arrays of Josephson junctions were presented
in Refs.\onlinecite{pfeiffer06,pfeiffer08}. Their experimental data demonstrate a series
of resonances in the current voltage characteristics of the array and were
analyzed using the discrete sine-Gordon model and an extension of this model
which includes a capacitive interaction between neighboring Josephson junctions.

To simulate the CVCs of intrinsic Josephson junctions in high
temperature superconductors we investigate the phase dynamics within
the framework of the CCJJ+DC model.
\cite{ryndyk-prl98,machida00,sms-physC06} In this model the dynamical equations are given by,
\begin{subequations}
\label{system}
\begin{eqnarray}
\frac{{\textrm d}V_{l}}{{\textrm d}t} = I+I_n -\sin \varphi_{l} -\beta\frac{{\textrm d}\varphi_{l}}{{\textrm d}t}, \label{eq1a} \\
\frac{{\textrm d}\varphi_{l}}{{\textrm d}t}= V_l - \alpha (V_{l+1}+ V_{l-1} - 2V_{l}). \label{eq1b}
\end{eqnarray}
\end{subequations}
Here $V_{l} \equiv V_{l+1,l}$ is the voltage difference between the $(l+1)$th and $l$th S-layers, and
the gauge-invariant phase difference $\varphi_{l}$ between
adjacent S-layers is given by $\varphi_{l}(t)=
\theta_{l+1}(t)-\theta_{l}(t)-\frac{2e}{\hbar c}\int^{(l+1)d}_{ld+s}{\rm
  d}z A_{z}(z,t)$, where $\theta_{l}$ is the phase of the order
parameter (i.e. phase of the macroscopic wave function of the
superconductor) in the $l$th S-layer of thickness $s$, $A_z$ is the vector potential in
the barrier of thickness $D$, and $d=s+D$ is the period of the lattice. In the above equations, $\alpha$ and
$\beta$ are the coupling constant and dissipation parameter,
respectively. Time $t$ is normalized to the inverse plasma frequency
$\omega^{-1}_p$ (with $\omega_p^2=2eI_c/\hbar C$, and $C$ denoting the
junction capacitance), the voltage differences $V_l$ are normalized to $V_0=\hbar\omega_p/(2e)$, and the bias current $I$ and added noise current $I_n$ are normalized to the critical current $I_c$. The role of the added noise current ($I_n \sim 10^{-8}$) has been discussed previously.\cite{sm-sust07} We solve this system of dynamical equations for stacks with various numbers ($N$) of JJs.

The system of Eqs. (\ref{system}) can also be written in the form
\begin{eqnarray}
\frac{{\rm d}^2\varphi_l}{{\rm d} t^2} = \sum\sb{l'}\,A_{ll'}\left[
I + I_n -\sin\varphi_{l'}-\beta \frac{{\rm d}
\varphi_{l'}}{{\rm d} t} \right] \label{model}
\end{eqnarray}
where, for nonperiodic boundary conditions (BCs),
the matrix $A$ has the form
\begin{eqnarray}
A=
\begin{pmatrix}
1+\alpha G&-\alpha&0&...&&&\\
-\alpha&1+2\alpha&-\alpha&0&...&&\\
0&-\alpha&1+2\alpha&-\alpha&0&...&\\
...&...&...&...&...&...&...\\
&&&...&0&-\alpha&1+\alpha G
\end{pmatrix}
\label{A-matrix}
\end{eqnarray}
Here $l$ and $l'$ run over the $N$ junctions, $\beta_c = \omega_p ^2 R^2 C^2$ is the McCumber
parameter, and $G=1+\gamma$, $\gamma=s/s_0=s/s_N$, where $s$, $s_0$ and $s_N$ are the thicknesses of the middle, first, and last $S$ layers, respectively.
According to the proximity effect, the thicknesses of the first and last layers are assumed to be thicker than the middle layers inside the stack. So, for the nonperiodic boundary BCs, the equations for the
first and last layers in Eq. (\ref{model}) are different from those of the middle layers.\cite{koyama96,matsumoto99} For periodic BCs the matrix $A$ has the form
\begin{eqnarray}
A=
\begin{pmatrix}
1+2\alpha&-\alpha&0&...&&&-\alpha\\
-\alpha&1+2\alpha&-\alpha&0&...&&\\
0&-\alpha&1+2\alpha&-\alpha&0&...&\\
...&...&...&...&...&...&...\\
-\alpha&&&...&0&-\alpha&1+2\alpha
\end{pmatrix}
\label{A-matrix-p}
\end{eqnarray}

Using the Maxwell equation  $\varepsilon\varepsilon_0 \textrm{div} {\vec E} = Q$, where $\varepsilon$ is the dielectric constant of the insulating layers and $\varepsilon_0$ is the permittivity of free space, we express the charge density in the $l$th S-layer as
\begin{equation}
Q_l=Q_0 \alpha (V_{l+1}-V_{l}), \label{q}
\end{equation}
where $Q_0 = \varepsilon \varepsilon _0 V_0/r_D^2$ and $r_D$ is the Debye screening length.
In all our calculations the CVCs and time dependence of the charge
oscillations in the S-layers are simulated at $\alpha = 1$. The system not very sensitive to the value of the coupling parameter. Further details concerning numerical simulations of the CVCs and the electric charge can be found in Refs. \onlinecite{smp-prb07,shk-prb09,matsumoto99}.

The usual test for chaos in a system is through the calculation of the largest Lyapunov exponent (LE).\cite{chaosb}  The general
idea is to follow two nearby trajectories and to calculate their average logarithmic rate of separation.  Whenever the two trajectories get too far apart, one of them has to be moved back to the vicinity of the other, {\em along the line of separation}. A conservative procedure is to do this at each iteration.

In this paper we solve the system of Eqs. (\ref{system}), using a fourth-order Runge-Kutta method, to find the $V_l$  and $\varphi_l$.
To this end we choose the time domain $T_f$ and time step $T_p = T_{f}/m$, such that $t_j= jT_p$, for $j=0,\ldots,m$.
In order to calculate the largest LE, at fixed current
$I$, we first integrate the system for a time $T_i$, until all transients have decayed. Then, starting from an arbitrary small initial separation vector $ {\vec d}_j \equiv (\Delta V(t_j), \Delta \varphi (t_j))$,
we advance both systems by one time step and calculate the new
separation
$ {\vec d}_{j+1} \equiv ( \Delta V(t_{j+1}), \Delta \varphi (t_{j+1}))$,
where $\Delta \varphi_l(t_i)=\varphi'_l(t_i)-\varphi_l(t_i)$ and $\Delta V_l(t_i) = V'_l(t_i)-V_l(t_i)$. The Lyapunov exponent for this step is defined by
\begin{equation}
LE_i=\frac{1}{T_p}\ln \frac{\| {\vec d}_{j+1} \|}{\| {\vec d}_{j} \|},
\end{equation}
where
$ \| {\vec d}_j \| = [\sum_{l=1}^{N}(\Delta \varphi_l(t_j))^2+(\Delta V_l(t_j))^2]^{1/2}$. In order to avoid cumulative numerical round-off errors, due to the expected exponential changes in separation, the separation vector is rescaled after each step in time, i.e. we set
${\vec d}_{j+1} \equiv ( \| {\vec d}_{j} \| / \| {\vec d}_{j+1} \| ){\vec d}_{j+1} $, before advancing to the next step. This
procedure is iterated over the rest of the time domain at fixed current. Finally we take the average of all the $LE_j$ to find the Lyapunov exponent
\begin{equation}
LE(I)=\frac{1}{m-k}\sum_{j=k}^{m-1}LE_j  \, . \label{LE}
\end{equation}
where $t_k=kT_p=T_i$. By changing the current in small steps (${\rm d}I$), and repeating the above procedure for each value of current, the $LE$ is obtained as a function of current. Typically, except where noted, we have used $T_i=50$, $T_f=25000$, $m= 5\times10^5$, and ${\rm d}I \le 5 \times 10^{-4}$. Additional details concerning the LE calculation can be found in Refs. \onlinecite{chaosb,chaos1,chaos2}.

\section{Indications of chaos in the BPR for a system of coupled JJ\lowercase{s}}
The CVCs of coupled JJs has a multiple
branch structure. Here we concentrate on the outermost branch
only. The system of coupled JJs exhibits a parametric resonance which
is reflected in the CVC as a breakpoint. In stacks with odd numbers of
junctions a breakpoint region, with a fine structure corresponding to
the width of the parametric resonance, is observed. A study of the
fine structure in the BPR gives one enough reason to look for chaos in
the dynamics of a stack. Its manifestation depends strongly on
the parameters of the system.  Here we present some examples and
methods, which might be used for demonstration and investigation of
the chaotic features of coupled JJs.

\begin{figure}[h!]
\includegraphics[width=0.45\textwidth]{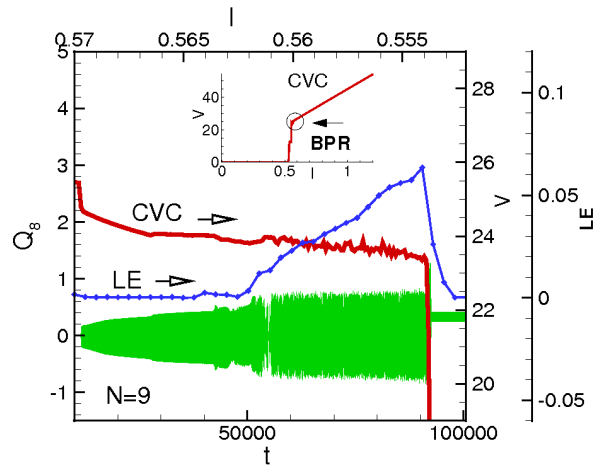}
\caption{(Color online) Lyapunov exponent and CVC, as functions of current, with the charge oscillation in $Q_8$ (the charge in the 8th S-layer) as a function of time. Inset demonstrates the position of the BP on the outermost branch of the CVCs.}
\label{fig1}
\end{figure}

First, we consider a stack of 9 JJs using periodic BCs with $\beta = 0.2$. In Fig. \ref{fig1}, the Lyapunov exponent and outermost branch of
CVCs, as functions of current, are compared with the charge
oscillation in the 8th S-layer, as a function of time and current.

To calculate the voltages $V_{l}(I)$ at each point of CVC
(i.e. for each value of $I$), we simulate the dynamics of the phases
$\varphi_{l}(t)$ by solving the system of equations  (\ref{system}) using the fourth-order
Runge-Kutta method with a time step $T_{p}$. After simulation of the phase
dynamics we calculate the dc voltages on each junction by averaging over a
chosen time domain. After completing the voltage averaging for the fixed bias
current $I$, the bias current is increased or decreased by a small amount $dI$
(bias current step) in order to calculate the voltages in all junctions at the
next point of the CVC. We use the distribution of phases and their derivatives
at the end of the time domain of previous point as the initial distribution
for calculating the next point. In this manner the dependence of charge on
current `includes' the time dependence of each current step.

In the inset we illustrate the position of the BPR on the outermost branch of the CVCs.
As one can observe, there are two distinct regions: $LE=0$ and $LE>0$. The case $LE>0$ strictly
indicates that the system is chaotic over the approximate current interval of $ (0.5620,0.5536)$, while $LE=0$ indicates marginal stability for the system elsewhere. Both the CVCs and charge oscillations show chaotic features which coincide with $LE>0$.

\begin{figure}[h!]
\includegraphics[width=0.45\textwidth]{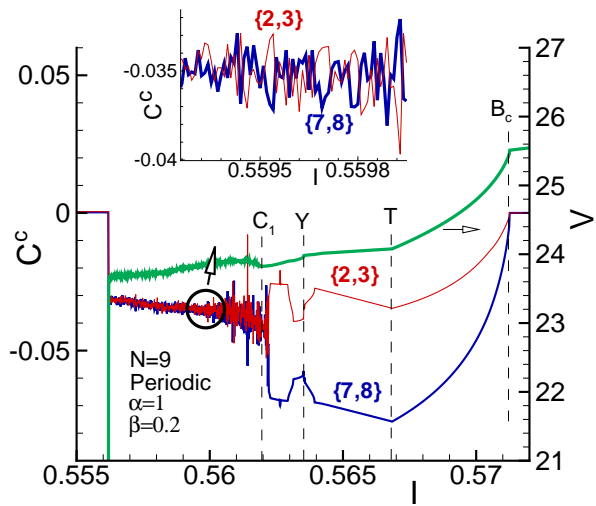}
\caption{(Color online)  The charge correlation functions $C^{c}_{l,l+1}=\left< Q_l(t) Q_{l+1}(t) \right>$ for $l=2$
(labeled by {2,3}) and $l=7$ (labeled {7,8}) for the stack with nine IJJ as a function of the bias
current. CVC is also shown (and labeled by an arrow). A chaotic part of the BPR (encircled) is enlarged in the inset.}
\label{fig2}
\end{figure}

Second, we study the correlations in superconducting currents in the neighboring Josephson junctions, and charge correlations in neighboring superconducting layers. This allows us to further demonstrate  the observed chaotic features in CVC. The  charge-charge correlation functions  for neighboring layers are given by
\begin{eqnarray}
C^{c}_{l,l+1}(I) & = & \left< Q_l(t)Q_{l+1}(t) \right> \label{cfeqn} \\
& = & \lim_{T_f\rightarrow\infty}\frac{1}{T_f-T_i}\int_{T_i}^{T_f}{Q_l(t)Q_{l+1}(t){\rm d} t}, \nonumber
\end{eqnarray}
where $l$ is the layer number and $T_i$ is the initial time for the averaging procedure. Figure~\ref{fig2} presents
the results of calculations for the cases $l=2$ and $l=7$, for the same stack of nine JJs. The negative sign of $C^{c}_{l,l+1}$ is due to the fact that a LPW with wave number $k = 8\pi/(9d)$ is created. This wave number is close to the $\pi$-mode, for which the charge on neighboring layers differs in sign, and hence the
product of positive and negative charges produces a negative sign (cf. Sec. IV).
Here the expected feature of the correlation functions in the chaotic region is confirmed: {\em at the transition to chaotic behavior (point $C_1$),
the values of all correlation functions approach each other, i.e.  $\left< Q_l(t)Q_{l+1}(t) \right>
=  \left< Q_l(t) \right> \left<Q_{l+1}(t) \right>$. Thus, within the chaotic region, all correlations are lost.}
To demonstrate the character of the correlation function, we enlarged a region in the chaotic part of the BPR in the inset to Fig. \ref{fig2}. We see that both correlation functions $C_{2,3}$ and $C_{7,8}$
show a similar variation with current, in agreement with the Lyapunov exponent.

\begin{figure}[h!]
\includegraphics[width=0.45\textwidth]{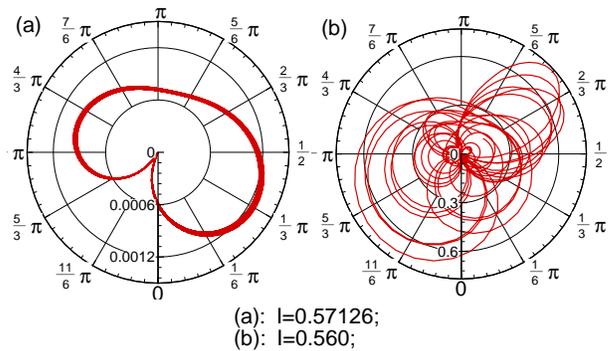}
\caption{\label{fig3}(Color online) The phase-charge diagram. It shows the absolute value of the charge on the first superconducting layer (radial coordinate), versus the phase difference across the first junction at two values of bias current: (a) $I=0.57126$ and
(b) $I=0.56000$.}
\end{figure}

Third, manifestations of the chaotic behavior can also be demonstrated by phase-charge diagrams. To do so, we record the time dependence of the phase difference across the chosen junction and electric charge in the chosen S-layer at a given bias current.
Figure~\ref{fig3} shows such a phase-charge diagram: the variation of the absolute value of charge in the first
superconducting layer $Q_1$ with the phase difference $\varphi_1$ across the first JJ of the stack.
This is a Lissajous figure, showing that the motion at $I=0.57126$ is qualitatively different from
that at $I=0.560$, thus providing additional evidence of chaos in this part of the CVCs.

\section{Effect of the number of junctions in the stack}
We next study the behavior of the stacks as a function of the number
of junctions. To clarify this effect, we first discuss the parametric resonance features in the stacks with even and odd numbers of JJs.

In the stacks with an \emph{even number of junctions} and periodic BCs, at $\beta=0.2$, a LPW with wave number $k=\pi$ ($\pi$-mode) is created, so that its wavelength $\lambda=2d$ satisfies the commensurability condition $L=n\lambda$.\cite{smp-prb07} Here $L=Nd$ is the length of the
stack, and $d$ is the period of the lattice.
In this case a fine structure in the BPR is absent and the transition from the outermost branch to
another one is observed after a sharp (exponential) increase of charge in S-layers. Also, the value of the breakpoint
current is the same for all stacks with an even number of junctions. The sharp increase of charge in the S-layers happens in a very narrow interval of the bias current. For example, in a simulation of the CVC in a stack with $N=10$ and a current step of $10^{-4}$, the interval is (0.5771, 0.5767) in which the increase of charge is of about 7 orders of
magnitude.

In the stacks with an \emph{odd number of junctions} the wave number of the LPW is $k=(N-1)\pi/(N d)$ which approaches  $k=\pi/d$ as $N$ increases. The value of the BP current is limited by its value at $k=\pi$, and the width of
the BPR decreases with increasing $N$.\cite{smp-prb07} With an odd number of junctions in the stack, the sharp increase of the charge accompanies the BP but then, before transition to another
branch, a wide BPR exists.\cite{sms-prb08} The origin of it is explained by the fact that the wavelength of LPW does not satisfy the commensurability condition. The fine structure is related to the character of the charge oscillations in the superconducting layers \cite{sms-prb08}. A part of BPR reflects the chaotic behavior of the system, which is the main subject of our paper. Next we look at how $N$ affects the structure of BPR. As mentioned above, the study of the correlation of charge on neighboring S-layers is a powerful tool to investigate the dynamics of charge in coupled JJs, within the BPR.\cite{shk-prb09} Here we use it to demonstrate the effect of the number of junctions on the structure of the BPR.

\begin{figure}[h!]
\includegraphics[height=90mm]{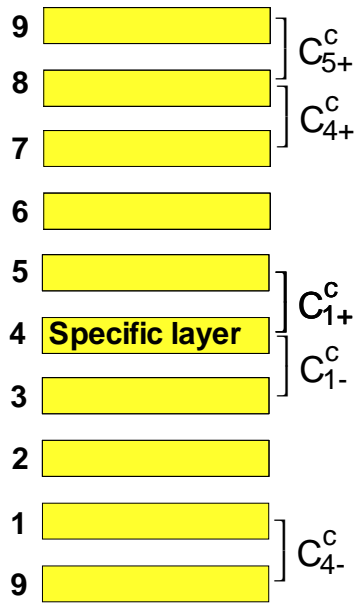}
\caption{(Color online) Formation of the charge correlation functions in a periodic stack of 9 IJJs. The numbers count the
layers in the stack.} \label{fig4}
\end{figure}

For the periodic BCs the number assigned to a junction is merely a label. However, in order to compare the correlation functions between stacks {\em with different numbers of junctions, in different simulations}, it is useful to use a specific layer as reference. In the parametric resonance region the phase dynamics in the coupled system of JJs
singles out such specific layers, namely, if a node of the charge standing wave coincides with the layer, then the charge on that layer is close to zero. Therefore we number the correlation functions according to their distance from such specific layers, as shown in Fig.~\ref{fig4}.
To reflect the symmetry of the correlation function relative to the chosen specific layer, we introduce the subscripts $+$ and $-$ on the correlation functions, as shown in Fig. \ref{fig4}.
In this case the specific layer is number 4, and we have $C^{c}_{1+}= C^{c}_{4,5}$  and $C^{c}_{1-}= C^{c}_{3,4}$ and so on.

\begin{figure}[h!]
\includegraphics[width=0.45\textwidth]{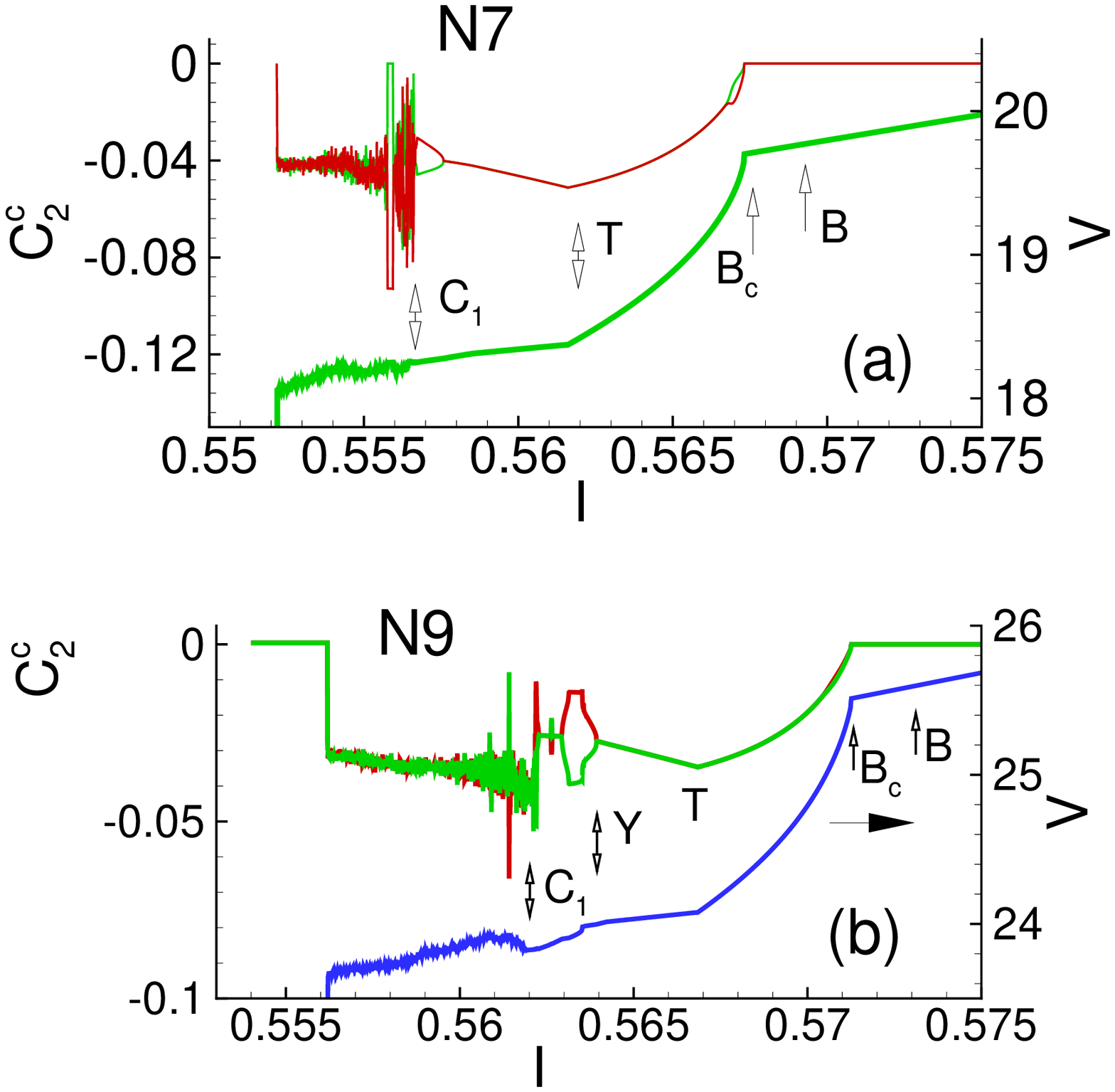}
\includegraphics[width=0.45\textwidth]{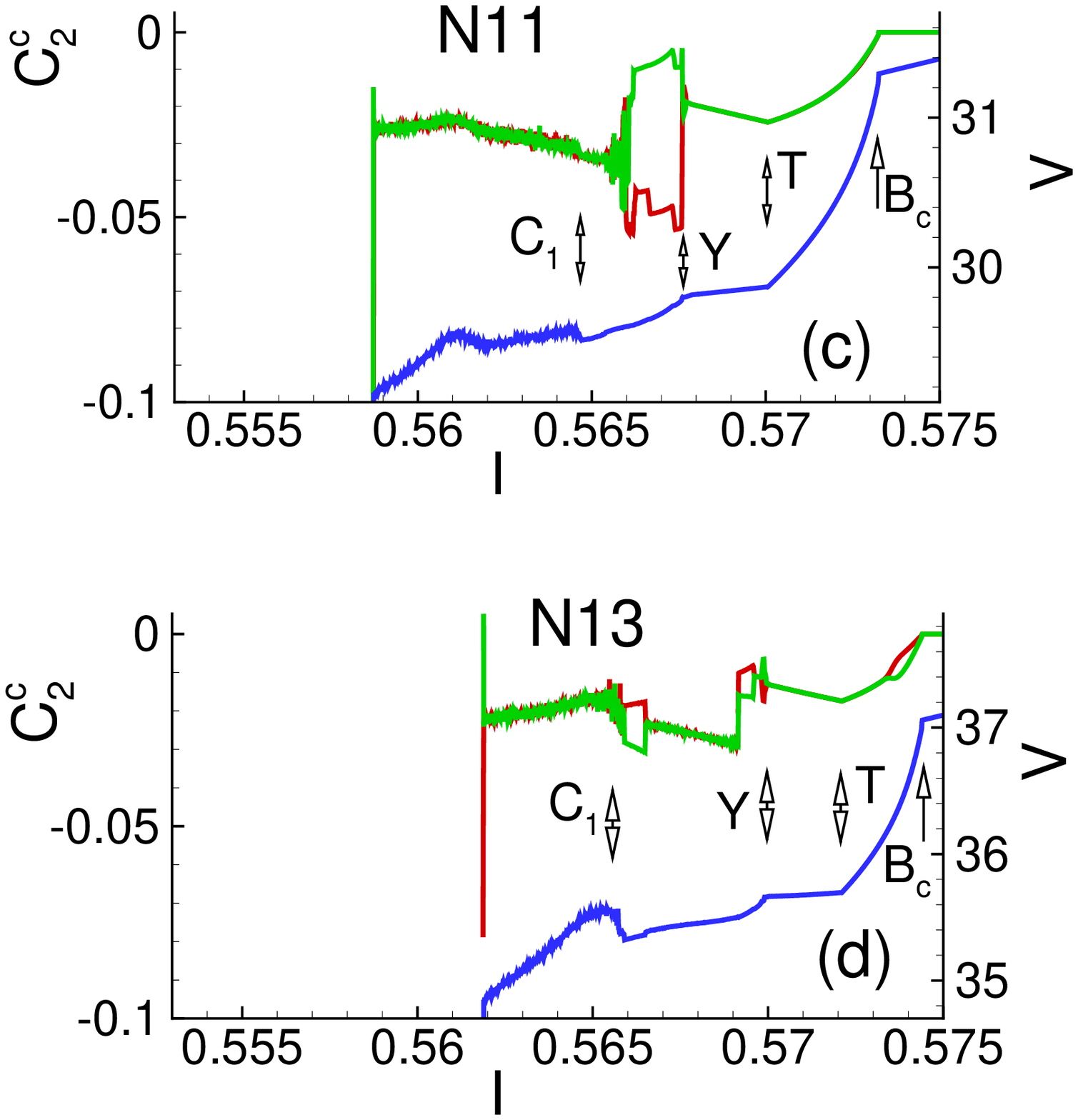}
\caption{(Color online) The CVC of the outermost branch in BPR together with correlation functions $C^c_{2+}$ and $C^c_{2-}$ for periodic
stacks with different numbers of IJJs: (a) $N=7$; (b) $N=9$;
(c) $N=11$; (d) $N=13$.
}  \label{fig5}
\end{figure}

Figure~\ref{fig5} shows the part of the outermost branch of CVC in the BPR together with two correlation functions
$C^c_{2+}$ and $C^c_{2-}$ for the stacks with 7, 9, 11 and 13 JJs with periodic BCs, at $\beta=0.2$.
In each case the part around $B_c$ corresponds to the maximal increase of charge in the S-layers (c.f. the beginning of charge dependence in Fig. 2) and is characterized by a LPW frequency $\omega_{LPW}$. To distinguish different parts of BPR, we introduce the notation: $B$ for the beginning of the parametric resonance, $B_c$ for the BP of the CVC, $T$ and $W$ to indicate the appearance of low frequency charge modulations, and $C_1$ to separate the chaotic part of the BPR. The loops that are seen in the correlation functions to the left of $B_c$ (very clearly manifested for $N=13$), correspond to the charge dynamics on the specific layer in the regular part of the BPR and were discussed in Ref.~\onlinecite{shk-prb09}.

In the part $T-Y$ a modulation of the charge oscillations with low frequency appears.\cite{sms-prb08} The part $Y-C_1$ is the transition region between regular and chaotic behavior where a second low frequency is observed. The features in the transition region $Y-C_1$ relate to the specific behavior of the LPW with $k=(N-1)\pi/(N d)$ in the stacks with different $N$. We found that the region after $C_1$  demonstrates the chaotic behavior of coupled
system of Josephson junctions.

From Fig.~\ref{fig5} one can draw the following conclusion: {\em The increase in $N$ increases the value of the breakpoint
  current, shifting the position of the BPR and decreasing the width
  of the BPR}.  As one can see, the increase in $N$
retains the main features of the BPR structure for the stacks with 7,
9, 11 and 13 junctions.  Simulation of the time dependence of the
charge in S-layers in these regions confirms the same character of the
charge oscillations for all these stacks.  The detailed features seen
in the fine structure of the correlation functions, for the stacks
with different numbers of junctions, and hence also different lengths,
are related to the specific behavior of LPWs with different wave
numbers. As will be discussed in Sec.~\ref{sec6}, transitions from chaotic to regular and back can be observed clearly in, for example, Fig.~\ref{fig5}(a), at $I \approx 0.556$.

\begin{figure}[h!]
\includegraphics[width=0.45\textwidth]{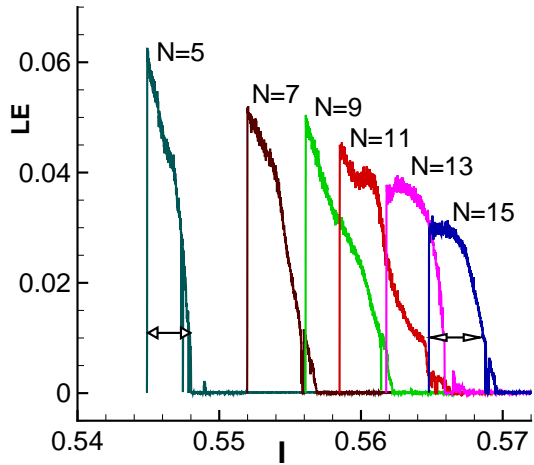}
\caption{(Color online) Maximal Lyapunov exponents for
periodic stacks with odd numbers of junctions $N$.
Double arrows emphasize the increase in width of the chaotic
region, relative to the BPR, as $N$ changes from $5$ to $15$.}
\label{fig6}
\end{figure}

Although the width of the BPR decreases with increasing $N$, the width corresponding to the chaotic behavior increases with increasing $N$. This fact is demonstrated in Fig. \ref{fig6}, where the LE as a function of $I$ is presented for stacks consisting of different numbers of junctions.
We see that the chaotic part of the CVC with five junctions is significantly smaller that for $15$ junctions. This fact has been stressed by the double arrows shown in Fig.~\ref{fig6}.

\section{Proximity effect}
Contact of the stack of superconducting layers with a normal metal
(electrodes) leads to the proximity effect: the superconducting
regions are expected to penetrate into the electrodes. Due to the proximity the effective thicknesses of the superconducting layers at the
edges of the stack, $s_0$ and $s_N$, are larger than the thicknesses, $s$, of the middle layers. As mentioned in connection with
Eq. (\ref{A-matrix}), this effect can be taken into account by
variation of the parameter $\gamma$ in the nonperiodic
BCs.\cite{matsumoto99} Therefore, in this section, we consider briefly the influence of $\gamma$ on the chaotic behavior of the CVC and correlation functions.

\begin{figure}[h!]
\includegraphics[width=0.45\textwidth]{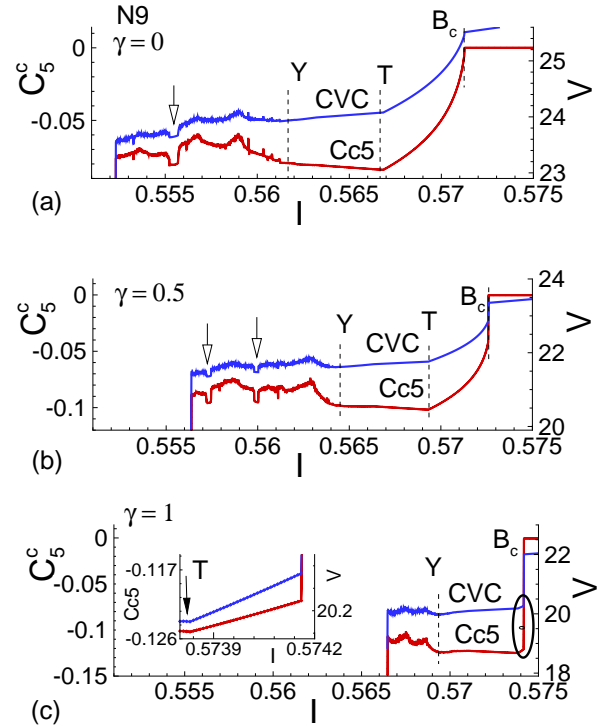}
\caption{(Color online) The CVC and correlation function ${C^c}_5$ in the  BPR for the  stack with 9 IJJ at  (a) $\gamma=0$; (b)  $\gamma=0.5$; (c) $\gamma=1$. Inset to (c) shows an enlarged view of the encircled area near the BP.}\label{fig7}
\end{figure}

The CVC and correlation function, $C^c_5=C^{c}_{5,6}(I) = \left<
Q_5(t)Q_{6}(t) \right>(I) $, at $\gamma = 0, 0.5 \mbox{ and } 1$ are
presented in Fig.~\ref{fig7} for a stack of 9 junctions. In this case,
the LPW with  $k=(N-1)\pi/N$ is created at the breakpoint for all
stacks, with an even and odd number of junctions and with $\beta=0.2$.\cite{smp-prb07} Figure~\ref{fig7}(a) shows the case
$\gamma=0$. As we see, the main features of the CVC and the correlation functions are coincident. Within the chaotic region there appears
to be some regions with regular oscillations, marked by
 arrows. Figure~\ref{fig7}(b) clearly shows two such regions in CVC and in the correlation function at $\gamma=0.5$.
With increasing $\gamma$ the width of the BPR is decreased, especially
the $B_c-T$ region, which becomes very small by the time $\gamma=1$. An enlarged view of this small region is shown in the inset to Fig.~\ref{fig7}(c).

\begin{figure}[h!]
\includegraphics[width=0.45\textwidth]{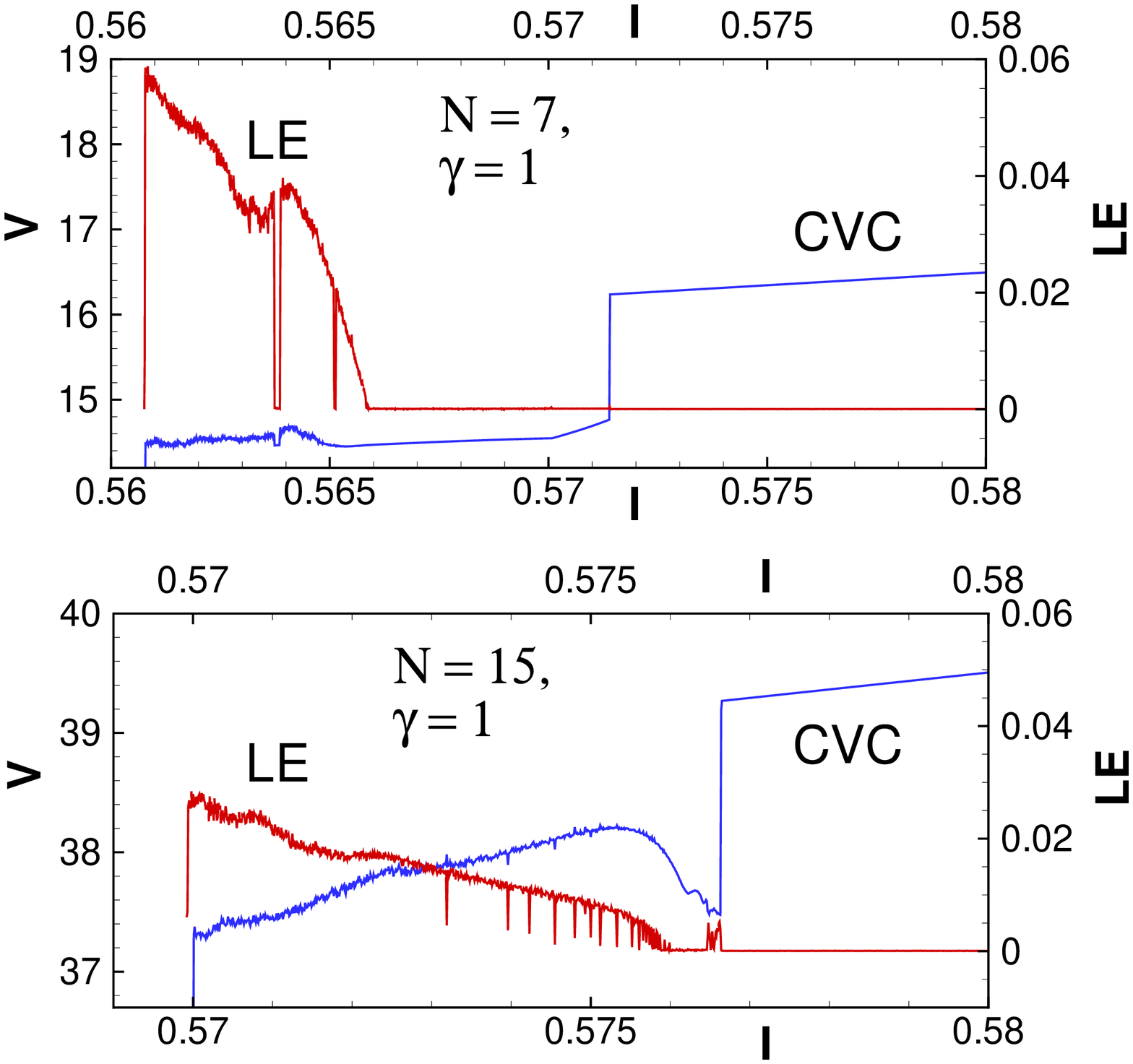}
\caption{(Color online) CVC of the outermost branch for stacks with (a) $N=7$ and (b) $N=15$ JJs together with the LE at $\gamma = 1$.}
\label{fig8}
\end{figure}

Taking into account the proximity effect further emphasizes the influence of the number of junctions in the stack. In particular, for $\gamma = 1$, the entire BPR becomes chaotic as $N$ increases beyond about 15. This result is presented in Fig.~\ref{fig8}, where the CVC of the outermost branch together with LE are compared for $N=7$ and $N=15$.
 We see that in the case $N=15$, the LE becomes positive practically
just after the breakpoint. In view of these results we predict that {\em the proximity effect will have a strong influence on the fine structure of the BPR in the CVC of real junctions}.

\section{Intermittency in CVC and correlation functions} \label{sec6}
\begin{figure}[h!]
\includegraphics[width=0.45\textwidth]{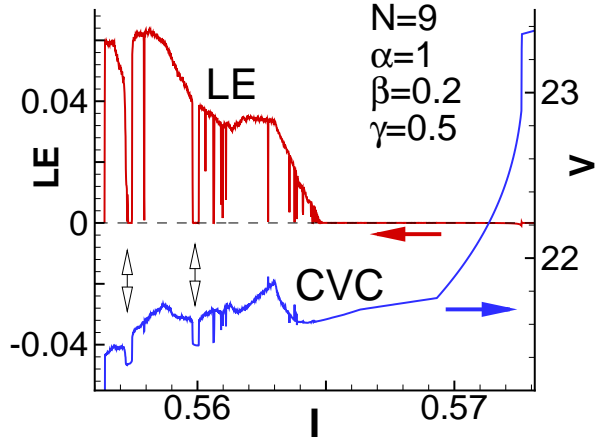}
\caption{(Color online) Intermittency is shown by windows of $LE=0$ inside the chaotic dynamics for a stack of 9 JJs with nonperiodic BCs. Arrows point to the respective scales for each curve and the dashed line shows the $LE=0$ axis. Double arrows show the correspondence between features of the CVC and the LE.} \label{fig9}
\end{figure}

As we have seen in Fig.~\ref{fig7}, there are some regular regions within the chaotic parts of the CVC and correlation functions, and so transitions exist from chaotic to regular behavior and back. This is known as intermittency in dynamical systems with
chaotic dynamics.\cite{roberto85,strogatz94} Now, for the first time, we demonstrate such intermittency for the resonance related chaos in systems of coupled JJs. Many such transitions can be seen in Fig.~\ref{fig9}, where we present results of a high precision calculation of the LE together with the CVC for a stack of 9 junctions, using nonperiodic BCs, at $\gamma=0.5$. In this calculation $T_i=50$, $T_f=250000$, $m=5 \times 10^{6}$  and ${\rm d}I = 10^{-5}$.
The regular regions seen in the CVC coinciding with $LE=0$. In particular the two largest intervals with $LE=0$ correspond to the current intervals $(0.5598,0.5600)$ and $(0.5573,0.5575)$, shown by double arrows.
To better understand the nature of these transitions, the charge time dependencies in these regions should be considered. In future work it would also be of interest to calculate the full spectrum of Lyapunov exponents, which would allow one to distinguish between chaotic behavior (one positive exponent) and hyperchaotic behavior (more than one positive exponent).

\begin{figure}[h!]
\includegraphics[width=0.45\textwidth]{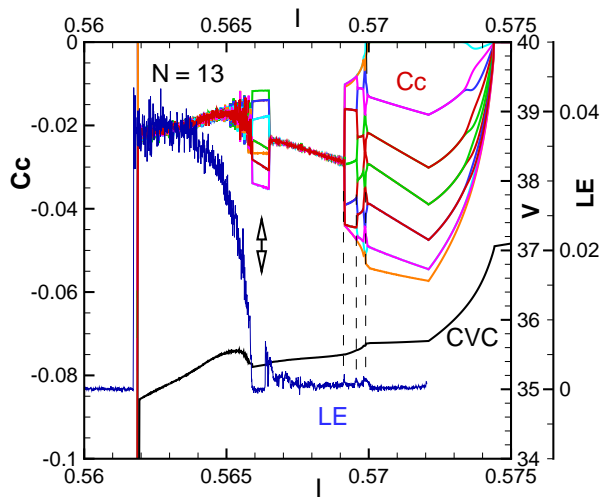}
\caption{(Color online) Demonstration of the intermittency within the interval of bias current $0.5659<I<0.5565$ (shown by a double arrow) for the charge-charge correlation functions (Cc) and LE for a periodic stack of 13 junctions.}
\label{fig10}
\end{figure}

To gain more information about the transitions we investigate the dependence of all the charge-charge correlation functions and the LE, on the bias current in the BPR. As the LE shows in Fig.~\ref{fig10}, the absence of the charge correlations in different S-layers is a signature of the chaotic behavior. Since we are interested in the chaotic features, we have not labeled each curve by its corresponding correlation function. In Fig.~\ref{fig10} we see a restoration of correlations in the middle of the chaotic region for a stack with 13 JJs. All presented characteristics (Cc, CVC and LE) reflect this transition from the chaotic behavior to regular and back.
To stress the agreement between the correlation functions and LE, vertical dashed lines have been drawn in Fig.~\ref{fig10}. We see that changes in the LE (small peaks on the LE curve) correspond exactly to changes in the charge correlations.

\begin{figure}[h!]
\includegraphics[width=0.45\textwidth]{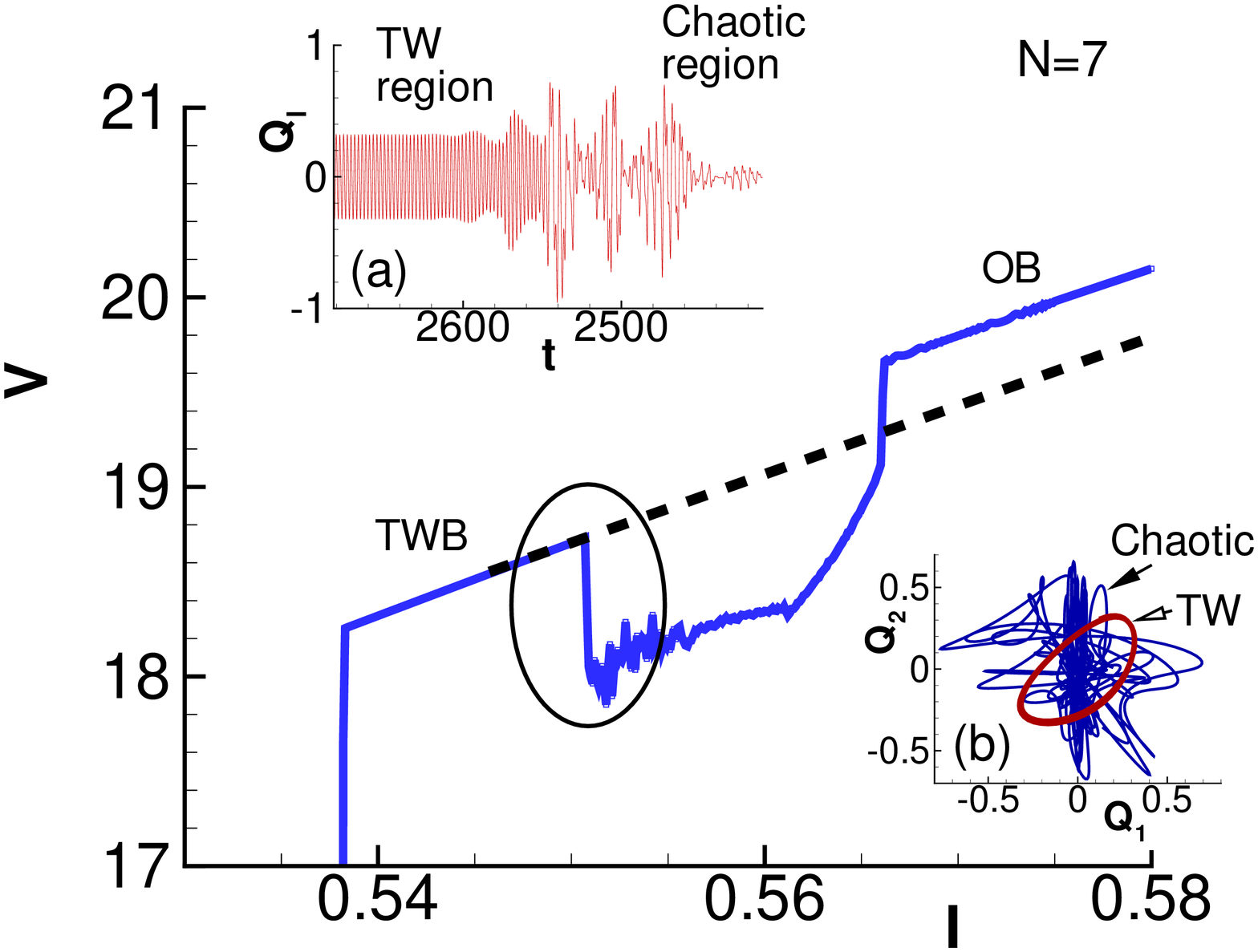}
\caption{(Color online) Transition in the outermost branch (OB) to the traveling wave branch (TWB) in the CVC of the stack with 7 JJ. Inset above the curve demonstrates charge vs. time for S-layer at the transition point. Inset below the curve shows the charge-charge diagram for the neighbor S-layers in the chaotic region of the OB and in TWB. \label{fig11}}
\end{figure}

We have characterized one such transition from the chaotic to regular behavior as corresponding to a charge traveling wave state.\cite{sh-zhetp-l-12} As shown in Fig.~\ref{fig11}, a transition from chaos to the traveling wave branch (TWB, dashed line) exists for the stack with 7 junctions, at $\beta=0.2$. This transition is encircled in the
figure.
We can see in the upper inset that the time dependence of charge in
this region demonstrates a changing of character of the charge
oscillations in S-layer. In the other inset (bottom right of
Fig.~\ref{fig11}) we have shown the Lissajous charge-charge diagram in
the chaotic state and in the TWB. The diagram demonstrate the
variation in time of the charges in two neighboring layers. The charge
in the first S-layer, $Q_1$, is plotted along the $x$-axis, and the
charge in the second S-layer, $Q_2$, is plotted along the $y$-axis. As
we can see, the Lissajous charge-charge diagram presents an open
trajectory for the chaotic region and a closed one for the TWB. Chaos
in systems of coupled JJs can also be transformed to regular behavior
by external irradiation. This is effected by adding a time dependent
current to $I$ to the right hand side of Eq. (1), through the
additional term, $A\sin (\omega_{R}t)$, where $A$ and $\omega_R$ are
the  amplitude and frequency of the microwave radiation.  This result
paves the way for experimental testing of the observed transitions
from chaotic to regular traveling wave behavior.  Detailed
investigations of the effects of radiation on IJJs will be made in a
future study.

\section{Linearized Equation: Parametric Resonance}
In this section we show that the linearized equation for the Fourier component $\delta_k$ of the phase difference  $\delta_l=\varphi_{l+1}-\varphi_{l}$ can determine the region of instability of the coupled JJ states, and that together with calculation of the Lyapunov exponent, in the  framework of the same equation, we can arrive at some important conclusions concerning the dynamics of this system. In particular,  \emph{there is a region of parameter values corresponding to the parametric resonance without transition to the chaotic state.} This conclusion is supported by the results of the investigation of the temporal dependence of the electric charge in S-layers and the correlation functions.

To investigate the instability region of coupled JJ states, corresponding to the outermost branch, we use the following linearized equation.\cite{sm-sust07,smp-prb07,sm-prl07} It can be obtained from Eq. (\ref{model}) as,
\begin{equation}
\ddot{\delta}_l+(1-\alpha\nabla^{(2)})(\cos(\varphi)\delta_l +\beta\dot{\delta}_l)=0,
\label{d-e-dp-l}
\end{equation}
by using the linear approximation
$\sin(\varphi_{l+1})- \sin(\varphi_{l}) \approx  \delta_l \cos(\varphi) $, where $\varphi\simeq\Omega t =
Vt/N$,  $\Omega=V/N$ is the Josephson frequency, and $V$ is the total voltage across the stack. Use of the Fourier expansion
$\delta_l(t)=\sum_k\delta_k e^{ikl}$, produces
\begin{eqnarray}
\ddot{\delta_k}+\beta(k)\dot{\delta_k}+ \cos(\Omega(k)t)\delta_k=0, \label{fur}
\end{eqnarray}
where  $t$ is normalized to $\omega^{-1}_{p}(k)$, with $\omega_p(k)=\omega_p C_{\alpha}$, $\beta(k)=\beta C_{\alpha}$,
$\Omega(k)=\Omega/C_{\alpha}$ and $ C_{\alpha} = \sqrt{1+2\alpha(1-\cos(k))}$. In Eq. (\ref{d-e-dp-l}), we have used the discrete Laplacian $\nabla^{(2)}f_l=f_{l+1}+f_{l-1}-2f_{l}$.

The linearized equation ~(\ref{fur}) demonstrates the instability region  on the  $\beta(k)-\Omega(k)$ diagram, which characterizes the parametric resonance in the coupled JJ.\cite{sm-prl07} This linear equation is similar
to a damped Mathieu type equation, with its periodic coefficient, demonstrating parametric instability (resonance).\cite{mathieu}

\begin{figure}[h!]
\includegraphics[width=0.45\textwidth]{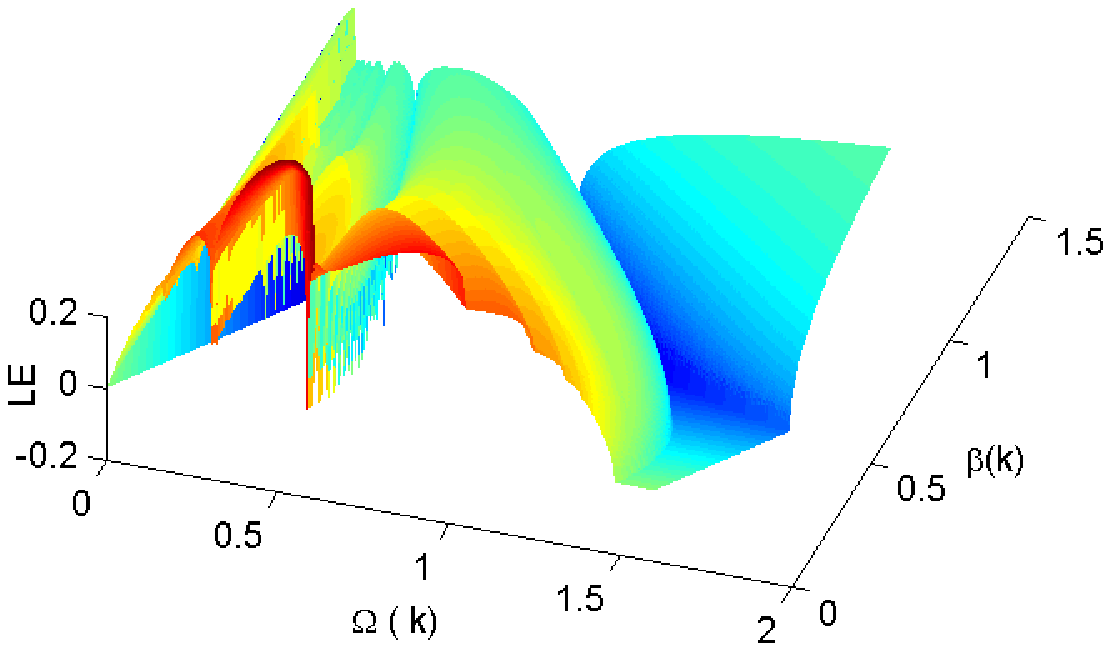}
\includegraphics[width=0.45\textwidth]{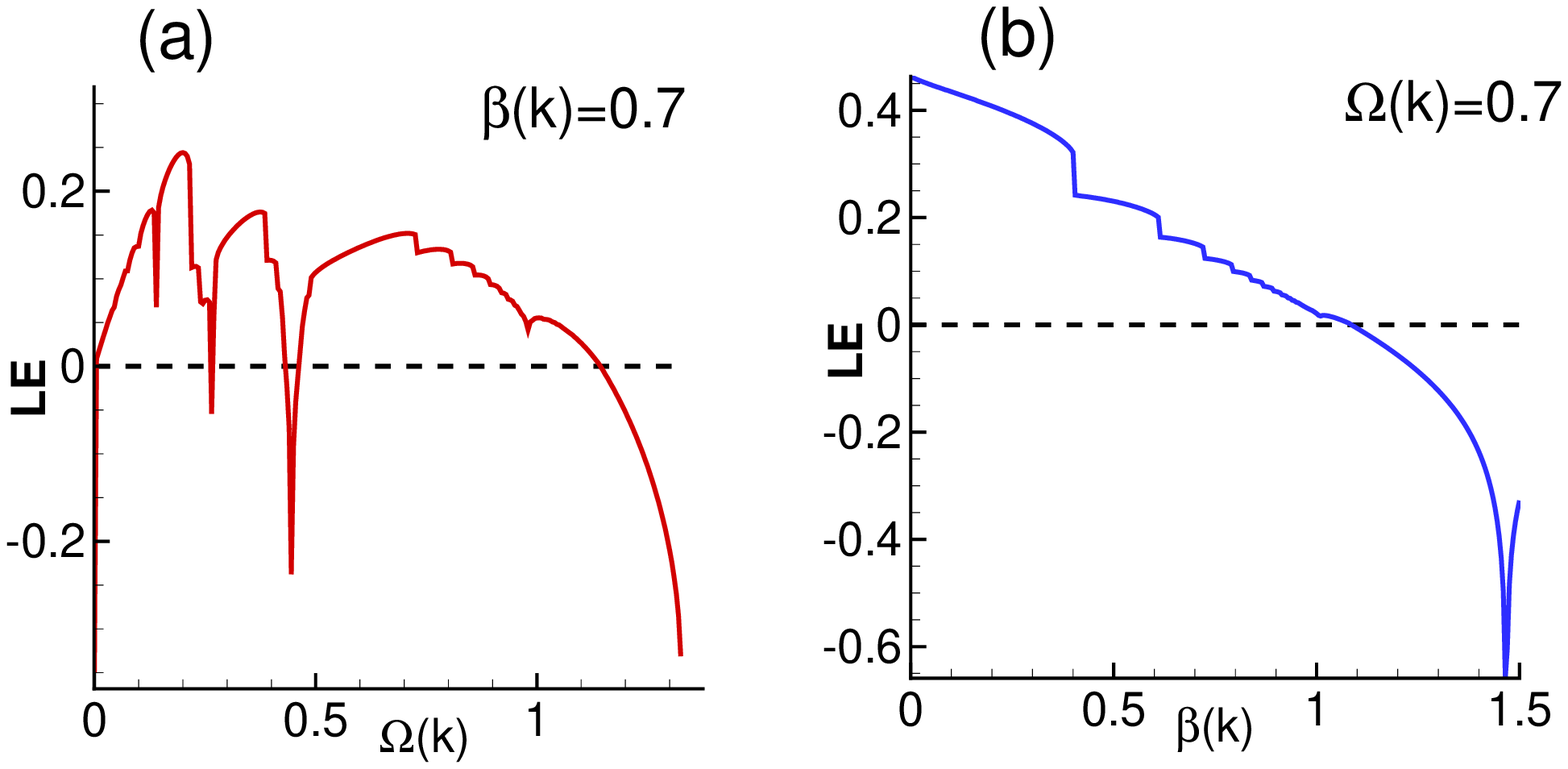}
\caption{\label{fig12} (Color online) Three dimensional picture of dependence of  Lyapunov exponent versus  $\Omega(k)$ and $\beta(k)$ given by equation (\ref{LE}). (a) The dependence of LE on $\Omega(k)$ at fixed $\beta(k)$; (b) The dependence of LE on $\beta(k)$ at fixed $\Omega(k)$.}
\end{figure}

\begin{figure}[h!]
\includegraphics[width=0.45\textwidth]{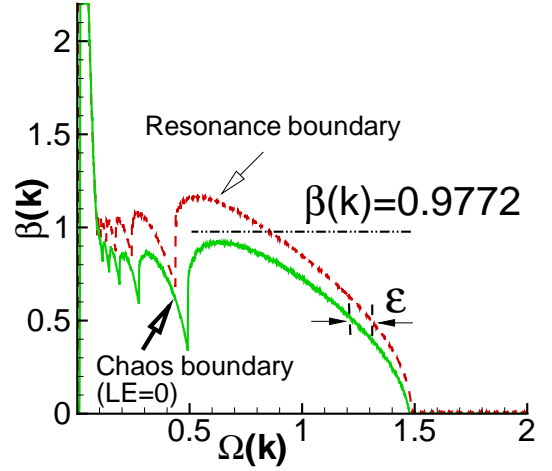}
\caption{\label{fig13} (Color online) Lyapunov exponent (solid) and resonance region (dashed) in $\Omega(k)$-$\beta(k)$ plane. $\varepsilon$ denotes the width of regular region in direction of $\Omega(k)$ axis. The horizontal double dashed dotted line corresponds to the value $\beta(k)=0.9772$. A system with this value of $\beta(k)$ doesn't touch the chaotic region. Below we demonstrate that chaotic part of BPR does not appear for this value of $\beta(k)$ (see Fig.~\ref{fig15}).    }
\end{figure}

We  calculate the LE for the dynamics dictated by the linearized equation (\ref{fur}), by the method described above, for the same system considered in Fig. \ref{fig1}. Three-dimensional results  are presented in  Fig.~\ref{fig12}. In our calculations we have taken $m=300000$, $d\Omega(k)=0.005$ and $d\beta(k)=0.005$. Below, we consider different cross sections of this figure. The LE has positive values at small values of $\Omega(k)$ and $\beta(k)$) and is negative for large $\Omega(k)$ and $\beta(k)$). We also show the boundary between these regimes by the curve denoting $LE=0$, in Fig~\ref{fig13}.

To clarify the results presented in Fig.~\ref{fig12}, we  demonstrate
two cross sections of it in the insets (a) and (b).  Inset (a) shows
the  LE at constant value of dissipation parameter $\beta(k)=0.7$ as a
function of $\Omega(k)$. We note that the $\Omega(k)$ is related to
the voltage by Josephson relation $\Omega(k)={V}/{(N C(k))}$,
i.e. $\Omega(k)$ is proportional to voltage.  The dashed line shows
$LE=0$. For large enough $\Omega(k)$, the LE is negative which points
to regular dynamics in the system. For lower values of $\Omega(k)$ the
behavior of the system is chaotic. Inset (b)  shows the LE at constant
value of voltage $\Omega(k)=0.7$ as a function of $\beta(k)$.

In Fig.~\ref{fig13} we show the cross section of Fig.~\ref{fig12} with
$LE=0$ plane (solid line) together with the instability region
(dashed line). Above the curve $LE=0$, the behavior of the system is
regular, and below it, chaotic. The border of the instability region
is related to the parametric resonance in the system. The region
between these two curves determines the regular behavior showing
parametric resonance.

Comparison of the Fig.~\ref{fig13} and Fig.~\ref{fig1} allows us to
estimate the width of the regular region at fixed $\beta(k)$, and
check the accuracy of linearization and approximations we  have used
to derive Eq. (\ref{fur}). From our definition of $\Omega(k)$, we have
$\Delta V=N C_k \varepsilon$, where $\varepsilon=\Delta \Omega(k) $
denotes the width of regular region in direction of $\Omega(k)$ axis
(see Fig.~\ref{fig13}).  In the case $k={8\pi}/{9}$
(the corresponding BPR in CVC for the 9 coupled JJs with $\beta=0.2$ is shown in Fig.~\ref{fig1}) we find $ C_{\alpha} =
\sqrt{1+2\alpha(1-\cos(k))}=2.2$. Then $\beta(k)=\beta
C_\alpha=0.44$. From Fig.~\ref{fig13} for this $\beta(k)$ we have
$\varepsilon=0.071$ and $\Delta V=N C_k \varepsilon\approx1.4$. As we
can estimate from  Fig.~\ref{fig1} the width of regular region is
$\Delta V\approx1.6$. So, the linearization and  approximations used
are  accurate   enough  and  there is a good agreement between results
presented in Figs.~\ref{fig13} and \ref{fig1}.

\section{Parametric resonance region without chaotic behavior: case $\beta=0.6$}

\begin{figure}[h!]
\includegraphics[width=0.45\textwidth]{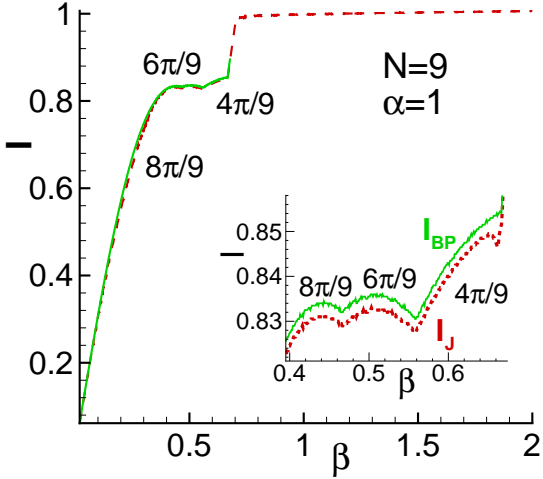}
\caption{\label{fig14} (Color online) Breakpoint current $I_{BP}$ and the jump current $I_J$ (where transition to another CVC branch occurs) versus $\beta$. Inset shows the enlarged part where changes of the wave number of LPW happen.}
\end{figure}

Let us now discuss in more detail the prediction which follows from Fig.~\ref{fig13}; namely, that there is an interval of $\beta(k)$ values for each fixed $\Omega(k)$ corresponding to the BPR without chaotic behavior. To find $\beta$ which corresponds to a $\beta(k)$ in the discussed interval, we should know the wave number of the created LPW.  We take an arbitrary value of $\beta(k)=0.9772$ in this interval. To determine the corresponding value of $\beta$ we first find the wave number of LPW which is created in this case at the resonance point. For this purpose we investigate, in Fig. \ref{fig14}, the $\beta$-dependence of the breakpoint current $I_{bp}$ (see details in Ref. \onlinecite{sm-prl07}) in Fig.~\ref{fig14}. Inset shows the enlarged part of the figure where changes of the wave number $k$ of the LPW happen. We see that the value of $\beta=0.6$ corresponds to $k=4\pi/9$.

We examine our idea at $\beta=0.6$ to check if for the stack with 9
coupled JJ a chaotic region is absent, and to see the character of
charge oscillations in the S-layers. According to the relation
$\beta(k)=\beta C_{\alpha}$, the value of $\beta=0.6$ follows from
that chosen in Fig.~\ref{fig13}, namely $\beta(k)=0.9772$ and the LPW
wave number $k={4\pi}/{9}$ because $ C_{\alpha} =
\sqrt{1+2\alpha(1-\cos(k))}=1.6287$.

\begin{figure}[h!]
\includegraphics[width=0.45\textwidth]{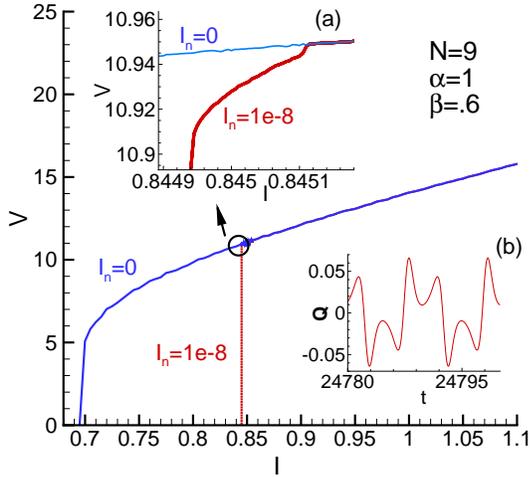}
\caption{\label{fig15} (Color online) CVC of the stack with N=9 Josephson junctions at $\beta=0.6$ and
periodic BC. Inset (a) demonstrates the enlarged part of BPR. Inset (b) shows electric charge on layers versus time.}
\end{figure}

In Fig.~\ref{fig15} we show the outermost branch in CVC of a stack of
$N=9$ coupled JJs with periodic BCs, at $\beta=0.6$.  Inset (a) to
this figure shows the enlarged BPR. In this inset we see that the
chaotic part of BPR does not appear. To stress the  absence of the
chaotic part in the BPR we also show the results for a system without
noise ($I_n=0$) in Fig.~\ref{fig15}.  In this case the system
behaves like a single JJ without a BPR in CVC. The
curves before the parametric resonance region coincide.
To see the dynamics of the electric charge in the S-layers, we show
the time variation of charge on one layer of a stack with 9 JJs with $\beta=0.6$ and periodic BCs in the inset (b) to Fig.~\ref{fig15}.  We see absolutely regular oscillations.

\begin{figure}[h!]
\includegraphics[width=0.45\textwidth]{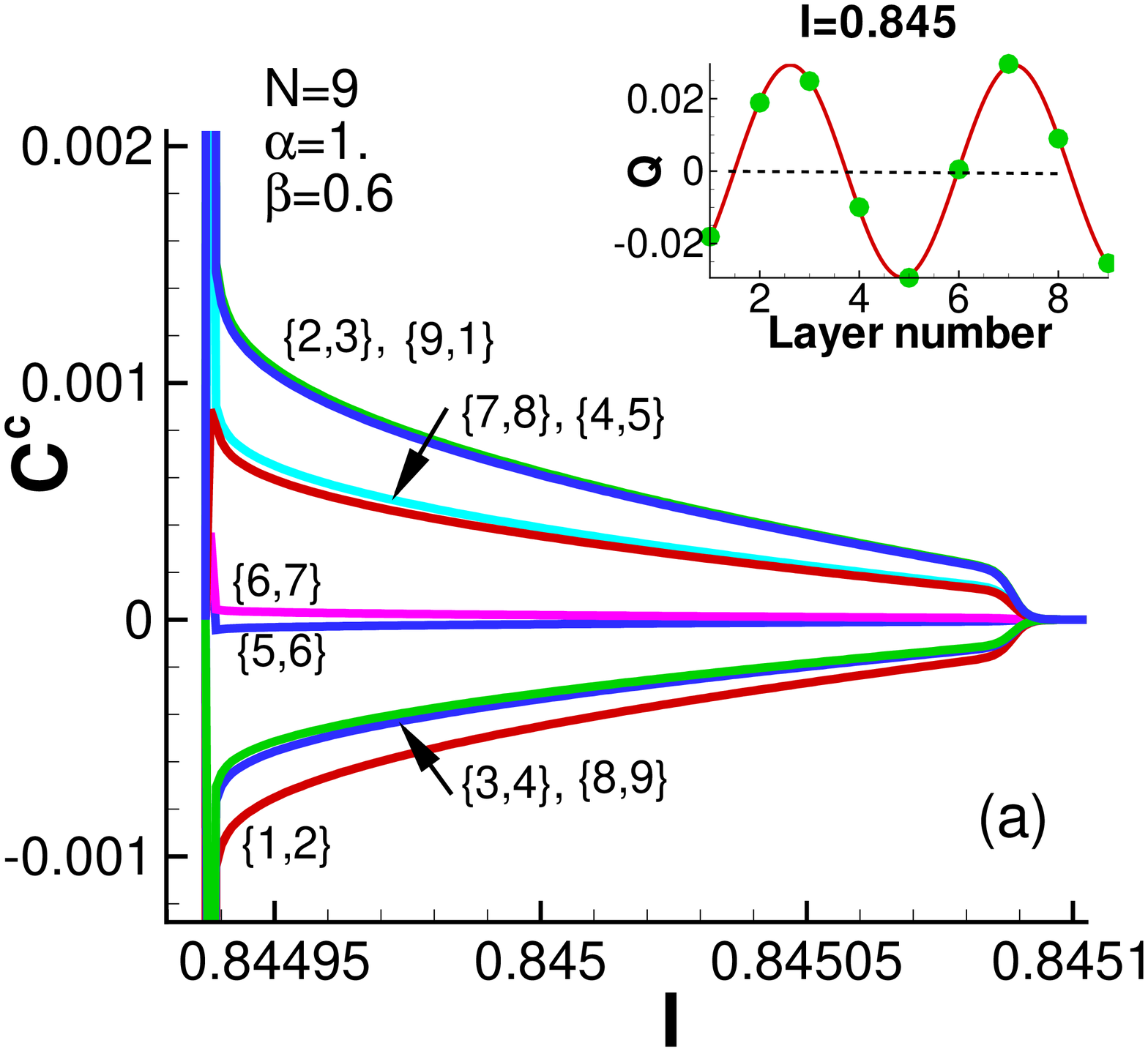}
\includegraphics[width=0.45\textwidth]{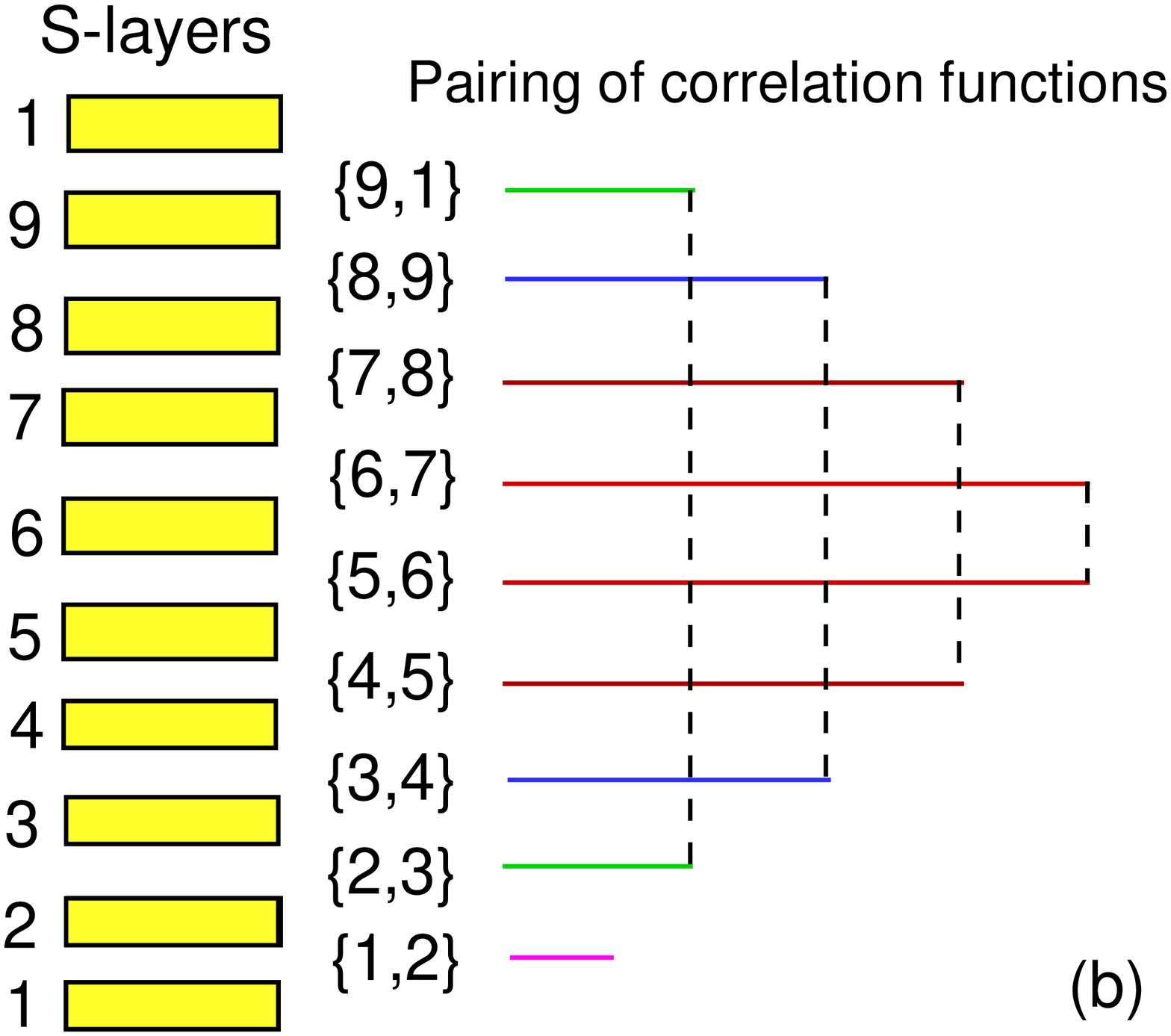}
\caption{\label{fig16} (Color online) (a) Charge-charge correlation functions versus bias current for the stack with 9 junctions for $\beta=0.6$.  Inset shows the charge distribution on S-layers at $I=0.845$. (b) Demonstration of correlation function pairing shown in (a).}
\end{figure}

Finally, we look at the charge correlations in neighboring S-layers
in a system without chaotic behavior in the parametric resonance region. As we know such a study is a powerful
tool for the investigation of the dynamics of coupled
JJs\cite{shk-prb09}.  Results of calculations are presented in
Fig. \ref{fig16} which demonstrates the charge-charge correlation
function (as defined in Sec. III) versus the bias current for
neighboring layers in the stack with 9 coupled JJs at $\beta=0.6$.
Here we are comparing the correlations within a specific stack, and so we use the usual notation for the correlation functions, i.e.
$C^{c}_{l,l+1} \equiv \{ l, l+1 \}$.
We see that the behavior of the correlation functions reflects the
behavior of the CVC in the BPR shown in Fig.~\ref{fig15}, and
corresponds to regular dynamics in the coupled JJs. The correlations
in the oscillations of the charge in S-layers persist up to the
transition to another branch in CVC. The inset to Fig.~\ref{fig16}(a)
shows the charge distribution on S-layers at $I=0.845$ and for the
wave number  $k={4 \pi}/{9}$. In Fig.~\ref{fig16}(b) we see that the
correlation functions form pairs corresponding to the standing
LPW. This pairing is demonstrated in Fig.~\ref{fig16}(b), from which
one can conclude that the 6th S-layer plays the role of the specific
layer. In this case the charge on the 6th layer is non-zero, because
the wavelength of the LPW is $\lambda=4.5d$ (not shown in
Fig.~\ref{fig16}(b)). Thus, the above analysis is a proof of concept
for the existence of parameters leading to a BPR without chaos.

\section{Conclusions}
Physical properties of intrinsic Josephson junctions in high
temperature superconductors continue to attract much attention  due to
the prospects they offer for superconducting electronics, in
particular, as high power sources of coherent electromagnetic
radiation in the THz region and  high precision voltage
standards. Applications would depend strongly on the chaotic
properties such systems. Observed radiation spectra from IJJs are very
complex and are temperature dependent. Transition of any junction in
the stack to the chaotic state would lead to a loss of synchronization
and coherent emission from the stack. The dissipation parameter
(McCumber parameter) is itself temperature dependent, and this has to
be taken into account when predicting the transition to the
chaotic state in real systems.

Here we have studied chaotic features of coupled intrinsic Josephson
junctions which might impact on their possible applications. Different
manifestations of the chaotic behavior were seen in the temporal
dependence of the electric charge in the superconducting layers,
phase-charge and charge-charge Lissajous diagrams, Lyapunov exponent
and correlation functions. We demonstrated that the
number of junctions in the stack and the boundary
conditions do influence on the chaotic part in
the breakpoint region. Experimental testing of the above features may therefore expedite progress in their potential applications. The chaotic part of BPR
increases with the number of junctions in the stack, relative to the
regular part of the BPR, at nonperiodic BCs. Transitions between
chaotic and regular behavior inside of the chaotic part of the
breakpoint region were seen. It is anticipated that future
developments in this direction of research may uncover new chaotic
aspects in coupled JJs.

Our analysis of the general features of the system has uncovered
the conditions and parameters that produce a breakpoint region without
chaos. We have shown that for high enough values of
dissipation parameter $\beta$ the chaotic part of breakpoint region
does not exist. This is reminiscent
of a single overdamped junction which lacks chaotic
dynamics.\cite{gitterman10} We also arrived at the important prediction
that the proximity effect should be expected to have
a strong influence on the fine structure of the breakpoint region.

In order to better understand the nature of the intermittency in
coupled JJs the charge time dependencies in these regions should be
considered in future work. It may also be of interest to calculate the
full spectrum of Lyapunov exponents, rather than only the maximal
exponent, in order to distinguish between chaotic and
(possibly) hyperchaotic behavior in these systems.

Presently there is renewed interest in phenomena related to the
switching from the outermost branch to inner branches and transitions
between inner branches. Recently, for example, powerful coherent THz
radiation was observed experimentally and was associated with a
``bump'' structure in  the same part of the CVC where the BPR was
found.\cite{kadowaki,benseman}  The recently observed broadly tunable
sub-terahertz emission from internal branches of the
CVCs\cite{kadowaki}  stresses the importance of further investigations
of the  chaos related to these inner branches.

\section{Acknowledgments}
Yu.M.S. thanks  Paul Seidel, Elena Zemlyanaya and Ilhom Rahmonov for stimulating discussions and V. V. Voronov, V. A. Osipov for supporting this work. Support from the South Africa-JINR collaboration is also acknowledged. M. R. Kolahchi acknowledges support from  the Institute for Advanced Studies in Basic Sciences, Zanjan, Iran. A.E.B. acknowledges that this work is based upon research supported by the National Research Foundation of South Africa.


\begin{thebibliography}{}
\bibitem{kleiner92} R. Kleiner, F. Steinmeyer, G. Kunkel, and P. Muller, Phys. Rev. Lett. {\bf68}, 2394 (1992).
\bibitem{kadowaki}  M. Tsujimoto {\em et al.}, Phys. Rev. Lett. {\bf 108}, 107006 (2012).
\bibitem{krasnov11} V. M. Krasnov, Phys. Rev. B {\bf 83}, 174517 (2011).
\bibitem{savelev10} S. Savel'ev, V. A. Yampol'skii, A. L. Rakhmanov, and F. Nori, Rep. Prog. Phys. {\bf73}, 026501 (2010).
\bibitem{ozyuzer} L. Ozyuzer {\em et al.}, Science {\bf 318}, 1291 (2007).
\bibitem{tachiki09} M. Tachiki, S. Fukuya, and T. Koyama, Phys. Rev. Lett. {\bf 102}, 127002 (2009).
\bibitem{wang10} H. B. Wang {\em et al.}, Phys. Rev. Lett. {\bf 105}, 057002 (2010).
\bibitem{lin11} S.-Z. Lin, X. Hu, and L. Bulaevskii, Phys. Rev. B {\bf 84}, 104501 (2011).
\bibitem{koshelev10} A. E. Koshelev, Phys. Rev. B {\bf 82}, 174512 (2010).
\bibitem{sg-prb11}Yu. M. Shukrinov and M. A. Gaafar, Phys. Rev. B {\bf 84}, 094514 (2011).
\bibitem{sm-prl07} Yu. M. Shukrinov, F. Mahfouzi, Phys. Rev. Lett. {\bf 98}, 157001 (2007).
\bibitem{sms-prb08} Yu. M. Shukrinov, F. Mahfouzi, M. Suzuki, Phys. Rev. B {\bf 78}, 134521 (2008).
\bibitem{sm-sust07} Yu. M. Shukrinov, F. Mahfouzi, Supercond. Sci. Technol., {\bf 19}, S38-S42 (2007).
\bibitem{iso-apl08} A. Irie, Yu. M. Shukrinov, and G. Oya, Appl. Phys. Lett. {\bf 93}, 152510 (2008).
\bibitem{machida99} M. Machida, T. Koyama, and M. Tachiki, Phys. Rev. Lett. {\bf 83}, 4618 1999.
\bibitem{watanabe95}  S. Watanabe, S. H. Strogatz {\em et al.}, Phys. Rev. Lett. {\bf 74}, 379 (1995).
\bibitem{benseman}T. M. Benseman {\em et al.}, Phys. Rev. B {\bf 84}, 064523 (2011).
\bibitem{ozyuzer12} F. Turkoglu {\em et al.}, Appl. Phys. Lett.; communicated.
\bibitem{Huberman80} B. A. Huberman, J. P. Crutchfield, and N. H. Packard, Appl. Phys. Lett. {\bf 37}, 750 (1980).
\bibitem{Kautz81} R. L. Kautz, J. Appl. Phys. {\bf 52}, 6241 (1981).
\bibitem{Kautz85a} R. L. Kautz and R. Monaco, J. Appl. Phys. {\bf 57}, 875 (1985).
\bibitem{Kautz85b} R. L. Kautz, J. Appl. Phys. {\bf 58}, 424 (1985).
\bibitem{pedersen} N. F. Pedersen and A. Davidson, Appl. Phys. Lett. {\bf 39}, 830 (1981).
\bibitem{irie03} A. Irie, Y. Kurosu, and G. Oya, IEEE Trans. Appl. Supercond. {\bf 13}, 908 (2003)
\bibitem{scherbel04} J. Scherbel {\em et al.}, Phys. Rev. B {\bf 70}, 104507 (2004).
\bibitem{tomas} T. Bohr, P. Bak, and M. H. Jensen, Phys. Rev. A {\bf 30}, 1970 (1984).
\bibitem{mogens} M. H. Jensen, P. Bak, and T. Bohr, Phys. Rev. A, {\bf 30}, 1960 (1984).
\bibitem{borcherds} P. H. Borcherds and G. P. McCauley, J. Phys. C {\bf 20}, 261 (1987).
\bibitem{benjacob} E. Ben-Jacob, I. Goldhirsch, Y. Imry, and S. Fishman, Phys. Rev. Lett. {\bf 49}, 1599 (1982).
\bibitem{osipov02} G. V. Osipov, A. S. Pikovsky and J. Kurths, Phys. Rev. Lett. {\bf 88}, 054102 (2002).
\bibitem{whan96}C. B. Whan and C. J. Lobb, Phys. Rev. E {\bf 53}, 405 (1996).
\bibitem{feng10} Y. L. Feng, X. H. Zhang, Z. G. Jiang and K. Shen, Int. J. Mod. Phys. B {\bf 24}, 5675 (2010).
\bibitem{ri11} I. Ri, Y. L. Feng,  Z. H. Yao, J. Fan, Chin. Phys. B {\bf 20}, 120504 (2011).
\bibitem{zhou09} T. G. Zhou, J. Mao, T. S. Liu, Y. Lai and S. L. Yan, Chin. Phys. Lett. {\bf 26}, 077401 (2009).
\bibitem{basler95} M. Basler, W. Krech and K. Y. Platov, Phys. Rev. B {\bf 52}, 7504 (1995).
\bibitem{hassel06} J. Hassel, L. Gronberg, P. Helisto and H. Seppa, Appl. Phys. Lett. {\bf 89}, 072503 (2006).
\bibitem{bhagavatula94} R. Bhagavatula, C. Ebner and C. Jayaprakash, Phys. Rev. B {\bf 50}, 9376 (1994).
\bibitem{chernikov94} A. A. Chernikov and G. Schmidt, Phys. Rev. E {\bf 50}, 3436 (1994).
\bibitem{koyama96} T. Koyama and M. Tachiki, Phys. Rev. B {\bf 54},  16183  (1996).
\bibitem{buckel} W. Buckel and R. Kleiner, {\em Superconductivity: Fundamentals and Applications} (Wiley-VCH, 2004).
\bibitem{machida00} M. Machida, T. Koyama, A. Tanaka and M. Tachiki, Physica C {\bf 330}, 85 (2000).
\bibitem{sms-physC06} Yu. M. Shukrinov, F. Mahfouzi, P. Seidel.  Physica C {\bf 449}, 62 (2006).
\bibitem{ryndyk-prl98} D. A. Ryndyk, Phys. Rev. Lett. {\bf 80}, 3376 (1998).
\bibitem{pfeiffer96} J. Pfeiffer {\em et al.}, Phys.Rev.Lett, {\bf96}, 034103 (2006).
\bibitem{pfeiffer98} J. Pfeiffer {\em et al.}, Phys.Rev.B, {\bf77}, 024511 (2008).
\bibitem{matsumoto99} H. Matsumoto, S. Sakamoto, F. Wajima, T. Koyama, and M. Machida, Phys. Rev. B {\bf 60}, 3666 (1999).
\bibitem{shk-prb09}Yu. M. Shukrinov, M. Hamdipour, M. R. Kolahchi, Phys. Rev. B {\bf 80}, 014512 (2009).
\bibitem{smp-prb07} Yu. M. Shukrinov, F. Mahfouzi, N. F. Pedersen, Phys. Rev. B {\bf 75}, 104508 (2007).
\bibitem{chaosb} J. C. Sprott, {\em Chaos and Time-Series Analysis} (Oxford University Press, 2003).
\bibitem{chaos1} H. Kantz, Phys. Lett. A {\bf 185}, 77 (1994).
\bibitem{chaos2} M. T. Rosenstein, J. J. Collins and C. J. De Luca, Physica D  {\bf 65}, 117 (1993).
\bibitem{roberto85} B. Roberto {\em et al.}, J. Phys.: Math. Gen. {\bf 18}, 2157 (1985).
\bibitem{strogatz94} S. H. Strogatz, {\em Nonlinear Dynamics and Chaos} (Addison-Wesley, 1994).
\bibitem{sh-zhetp-l-12} Yu. M. Shukrinov and M. Hamdipour, JETP Lett. {\bf 95}, 336 (2012).
\bibitem{mathieu} L. D. Landau, E. M. Lifshitz, {\em Mechanics} (Butterworth-Heinemann, Vol. 1, 3rd ed., 1976).
\bibitem{gitterman10} M. Gitterman, {\em The Chaotic Pendulum} (Wold Scientific, 2010).

\end{thebibliography}
\end{document}